\documentclass[twocolumn,showpacs,preprintnumbers,prl,aps]{revtex4-1}

\usepackage{graphicx}
\usepackage{dcolumn}
\usepackage{amsfonts,amsmath,amssymb,bm}
\usepackage{natbib}
\usepackage{hyperref}
\setcitestyle{super}

\begin{document}

\title{\bf {Electronic structure of Co doped ZnO from the \textit{GW} perspective}}
\date{\today} 
\author{I.~Abdolhosseini~Sarsari$^{a,b}$}
\author{C.~D.~Pemmaraju$^a$}
\author{Hadi~Salamati$^b$}
\author{S.~Sanvito$^a$}
\affiliation{a) School of Physics and CRANN, Trinity College, Dublin 2, Ireland\\
b)Department of Physics, Isfahan University of Technology, Isfahan, 84156-83111, Iran}

\newcommand{\etal}{{\em et al}}
\begin{abstract}

In transition metal doped ZnO, the energy position of dopant 3$d$ states relative to 
host conduction and valence bands is crucial in determining the possibilty of long range 
ferromagnetism. Density functional theory based estimates of the energy position of  
Co-3$d$ states in Co doped ZnO differ substantially depending upon the choice of 
exchange-correlation functional. In this work we investigate many-body $GW$ corrections
on top of DFT$+U$ and hybrid-DFT groundstates to provide a theoretical benchmark for 
the quasiparticle energies in wurtzite ZnO:Co. Both single shot $G_0W_0$ as well as partially 
self-consistent $GW_0$ wherein the wavefunctions are held fixed at the DFT level but the 
eigenvalues in G are iterated, are considered. The predicted energy position of the minority
spin Co-$t_2$ states is 3.0-3.6~eV above the ZnO conduction band minimum which is closer 
to hybrid-DFT based estimates.

\end{abstract}
\pacs{}
\keywords{}

\maketitle

\section{INTRODUCTION}
The quest for oxide dilute magnetic semi-conductors (DMS) exhibiting 
high Curie temperature has been ongoing for nearly a decade
driven by the prospect of realizing future spintronic materials
incorporating both semi-conducting and ferromagnetic properties~\cite{Dietl2000}.
ZnO, already of great technological relevance as a transparent conducting oxide
exhibiting a multitude of interesting optical and electrical properties, has
been widely studied as a potential DMS material following initial
reports of room-temperature ferromagnetism (RTF) in ZnO thin-films~\cite{Ueda2001}. 
Stabilizing high-T$_c$ ferromagnetism in ZnO in conjuction with its direct and wide band gap, 
large exciton binding energies, and large piezoelectric constants would lead to 
a truly multifunctional DMS~\cite{Janotti2009}. However, after several years of 
experimental and theoretical investigations a complete explanation of the ferromagnetism 
in Co doped ZnO (ZnO:Co) remains elusive~\cite{Dietl2010}. Recent experiments suggest a picture wherein 
ferromagnetism is absent in uniformly doped single-crystal ZnO:Co but emerges 
in highly defective poly-crystalline samples with extended defects such as grain-boundaries
playing a role~\cite{gamelin2008,hsu2007,Straumal2009}. 

Numerous theoretical efforts based on density functional theory (DFT)
have also investigated the microscopic origins of magnetic interactions between
Co ions in ZnO:Co~\cite{Spaldin2004,Patterson2006,Das2008,Walsh2008,Zunger2009}. 
In particular, a variety of beyond-LDA/GGA methodologies (indicated collectively as $b$-LDA) such as DFT+$U$~\cite{Walsh2008},
NLEP~\cite{Lany2008}, ASIC~\cite{Das2008} and hybrid-DFT~\cite{Patterson2006}
have been employed, to mitigate the severe band-gap underestimation in ZnO by semi-local functionals. 
The description of the ground-state electronic structure of ZnO:Co in the absence of additional charge doping defects
is similar in the different $b$-LDA approaches: Co$^{2+}$ ions doping the Zn site (Co$_\mathrm{Zn}$) in wurtzite ZnO are nominally 
in a $d^7$ valence state
and the approximately tetrahedral crystal field splits the Co-3$d$ states into a set of lower $e$ and higher $t_2$
like levels. The majority-spin $e$ and $t_2$ as well as the minority-spin $e$ states are filled while the
minority spin $t_2$ ($t_2^{\downarrow}$) states are empty leading to a net magnetic moment of 3$\mu$B per Co$_\mathrm{Zn}$. The energy position of the $t_2^{\downarrow}$ states relative to the host conduction band (CB) is however
crucial for ferromagnetism in ZnO:Co. Theoretical works in the literature~\cite{Das2008,Walsh2008,Zunger2009}
employing a variety of $b$-LDA approximations indicate that partial occupancy of the $t_2^{\downarrow}$ states
under extraneous electron doping is a minimum requirement for long-range FM interactions between Co ions in ZnO:Co.
Unfotunately, the different approaches differ substantially in their estimates for the position
of the $t_2^{\downarrow}$ states relative to the conduction band minimum (CBM), leading to 
different predictions for the feasibilty of RTF in ZnO:Co~\cite{Walsh2008,Zunger2009,Das2008,DasPRl2009,WalshPRL2009}. 
A scenario where the Co-$t_2^{\downarrow}$ states 
are located at or below the CBM~\cite{Walsh2008} would be conducive for ferromagnetism in ZnO:Co at 
modest n-doping and without the need for structural defects where as if the $t_2^{\downarrow}$ states were 
resonant well inside the conduction band, either larger n-doping~\cite{Zunger2009} or additional structural 
defects that lower the position of the $t_2^{\downarrow}$ states towards the CBM~\cite{Das2008} would be 
necessary to drive a FM state. The assumption underlying these predictions based on DFT is the interpretation 
of Kohn-Sham (KS) eigenvalues as approximate addition and removal energies that correspond to photoemission
spectra~(PES). While no formal justification exists for such an interpretation, $b$-LDA approaches are generally
designed to improve the agreement with experimental PES of either all or a subset of KS eigenvalues. In the 
absence to-date of direct observation of the empty Co-$t_2^{\downarrow}$ states by inverse photoemission experiments,
we seek to resolve the ambiguity in the theoretical description by directly 
calculating the quasiparticle (QP) spectrum of ZnO:Co within the $GW$ approximation 
($GWA$)~\cite{Hedin1969,Hybertsen1986,Shishkin2006}.

The many-body perturbation theory baseed $GWA$ is a popular approach for calculating the quasiparticle energies of solid-state
systems~\cite{Aryasetiawan1998}. Many-body effects in the electron-electron interaction that go beyond the mean-field picture are 
incorporated into the $GWA$ via the energy-dependent electron self-energy operator $\Sigma$
which is approximated as a product of the Green's function $G$ and the dynamically screened Coulomb interaction $W$. 
$W$ is in turn obatined by screening the bare-Coulomb interaction with the inverse frequency-dependent dielectric
matrix. $GW$ self-energy corrections are generally calculated on top of DFT independent-particle  wavefunctions and
eigenvalues and the resulting QP spectra are systematically improved towards direct/inverse 
photoemission spectra (PES). Different levels of self-consistency are possible within a perturbative GW scheme~\cite{Shishkin2007}
and in this work, we consider both single-shot $G_0W_0$ as well as partially self-consitent $GW_0$ wherein one iterates 
the eigenvalues in $G$ while keeping $W$ and the wavefunctions fixed at the DFT estimate. We find that irrespective
of the specific $b$-LDA starting point, quasi-particle energies of the Co-$t_2^{\downarrow}$ states are located well above 
the CBM of ZnO. Thus partial occupancy of these states is difficult to achieve for low electron doping concentrations.

\section{METHODS}
All the calculations presented in this work were carried out 
within the planewave based DFT framework as 
implemented in the standard VASP~\cite{VASP1,VASP2} package. 
A plane-wave kinetic energy cutoff of 300 eV and projector-augmented wave
(PAW) pseudopotentials \cite{Blochl1994} with the following valence-electron configurations
were employed: 3$d^{10}$4$s^2$ for Zn, 2$s^2$2$p^4$ for O and 3$d^8$4$s^1$ 
for Co. The pseudopotentials as well as pure DFT calculations in this work employed the 
PBE~\cite{PBE1996} exchange-correlation functional. 
Two different $b$-LDA approaches viz., PBE$+U$~\cite{Dudarev1998} and hybrid-DFT~\cite{Seidl1996,Heyd2003} 
were considered to provide starting points for subsequent $GW$ calculations. 
The PBE$+U$ calculations employed Hubbard parameters of
$U_\mathrm{Zn}$=7 eV, $U_\mathrm{Co}$=3 eV and $J_\mathrm{Co}$=1 eV, for the 3$d$ states
of Zn and Co respectively~\cite{Zunger2009}.
The bigger $U$ for Zn is attributed to a deeper and more localized semicore $d^{10}$ shell in 
Zn compared to Co.
Hybrid-DFT calculations employed the HSE03~\cite{Heyd2003} functional. 

For calculations on the wurtzite unit-cell of ZnO, the Brillouin zone was sampled 
using a 8$\times$8$\times$6 $\Gamma$ centered $k$-point mesh. Cobalt doped ZnO was modeled 
with a 32 atom orthorhombic supercell of wurtzite ZnO, in which one Zn site was substituted by Co.
This corresponds to a nominal Co doping concentation of $\sim$6.25\%.
In the supercell calculations, the Brillouin zone was sampled at 28 irreducible $k$-points in a $\Gamma$ centered mesh.
Structures were optimized using the HSE03 functional until residual forces were
smaller than 0.01 eV/\AA~(0.05 eV/\AA) in the primitive-cell (supercell). This led to the following unit-cell parameters for wurtzite ZnO: a=2.248~\AA,
c/a=1.61, u=0.380. Within $GW$ calculations, an energy cutoff of 150 eV was used for the response functions. 
A total of 240 and 1152 bands were employed in the primitive (wurtzite) cell and supercell calculations
respectively. Both single shot $G_0W_0$ as well as partially self-consistent $GW_0$ 
calculations were carried out on top of DFT based groudstate starting points. In the $GW_0$ calculations,
the eigenvalues in G were self-consistently updated four times while the orbitals were held 
fixed as obtained from DFT.

\section{RESULTS AND DISCUSSIONS}

\subsection{ZnO unitcell}
In order to set the framework for supercell defect calculations
on Co doped ZnO, we first investigate the eigenvalue spectrum 
of pure ZnO obtained both from DFT as well as $GW$ quasi-particle
corrections on top of DFT. 
In ZnO, the predominant character of the valence band maximum (VBM) 
and conduction band minimum (CBM) is O$_{2p}$ and Zn$_{4s}$ respectively.
The Zn$_{3d}$ states meanwhile, are fully occupied and are located several eV
below the VBM. The computed band gaps and Zn$_{3d}$ binding energies
are reported in table~\ref{Tab1}.  
ZnO is a prototypical case for extreme band-gap (E$_g$)
underestimation by semi-local exchange-correlation (XC) functionals. The predicted
band-gap from the PBE functional for instance is 0.78 eV while the experimental
gap is 3.44 eV~\cite{Kittel1986}. Some part of this 
band gap underestimation by semi-local functionals can be traced to the 
too low binding energy of cation $3d$ states and their concomitant
hybridization with anion 2$p$ states in the valence band. The average Zn$_{3d}$ 
binding energy (E$_{3d}$) from PBE is $\sim$5.1 eV compared to 7.5-8.81 eV in experiment~\cite{Ley1974}. 
Given a band-width of $\sim$5.5 eV for the O$_{2p}$ valence band, this leads to spurious 
Zn$_{3d}$-O$_{2p}$ hybridization which because of $pd$ replusion, 
pushes the O$_{2p}$ states higher in energy reducing E$_g$. 
This effect which can be traced to self-interaction errors,
is over and above the conventional DFT underestimation of band-gaps in 
semiconductors~\cite{Fuchs2007}.
\begin{table}[htb]
\caption{\label{Tab1} Calculated band gap (E$_g$) and average binding energy of Zn-$3d$ states~(E$_{3d}$)
in pure ZnO from different levels of theory are compared to experiment~\cite{Kittel1986,Ley1974}. 
All energy values are given in eV.
}
\begin{ruledtabular}
\centering
\begin{tabular}{c|c|c} 
Method          &E$_g$ (eV)               &E$_{3d}$ (eV) \\
\hline 
PBE                & 0.78                 & 5.15  \\
PBE+$G_0W_0$     & 2.27               & 6.05  \\
PBE+$GW_0$         & 2.68               & 6.39  \\
\hline
PBE+$U$             & 1.58            & 6.98   \\
PBE+$U$+$G_0W_0$  & 2.62            & 6.69   \\
PBE+$U$+$GW_0$      & 2.85            & 6.63   \\
\hline
HSE               & 2.24              & 6.01   \\
HSE+$G_0W_0$   & 3.14              & 6.64   \\
HSE+$GW_0$       & 3.31              & 6.81   \\
\hline
Exp.            & 3.44              & 7.5-8.81   \\ 
\end{tabular}
\end{ruledtabular}
\end{table}
A significant improvement in the 
value of E$_g$ can be obtained simply by correcting for the low binding energy
of Zn$_{3d}$ states. Accordingly, the inclusion of a Hubbard-$U$ correction on 
the Zn$_{3d}$ states within PBE+$U$ results in an  E$_{3d}$
of $\sim$7 eV while also improving E$_g$ to 1.58 eV. This nevertheless
still represents an almost 50\% underestimation of E$_g$. The description can 
be further improved by employing a hybrid-DFT functional such as HSE03. The inclusion of
non-local Fock exchange not only leads to a reduction in the self interaction error 
but to a large extent also restores the derivative discontinuity in the XC functional 
within a generalized Kohn-Sham ($GKS$) scheme~\cite{Seidl1996}, further improving the band-gap. 
Thus HSE03 predicts values of E$_g$ and E$_{3d}$ at 2.24 eV and 6 eV respectively. 
\begin{figure*}[htbp]
\includegraphics[width=\textwidth]{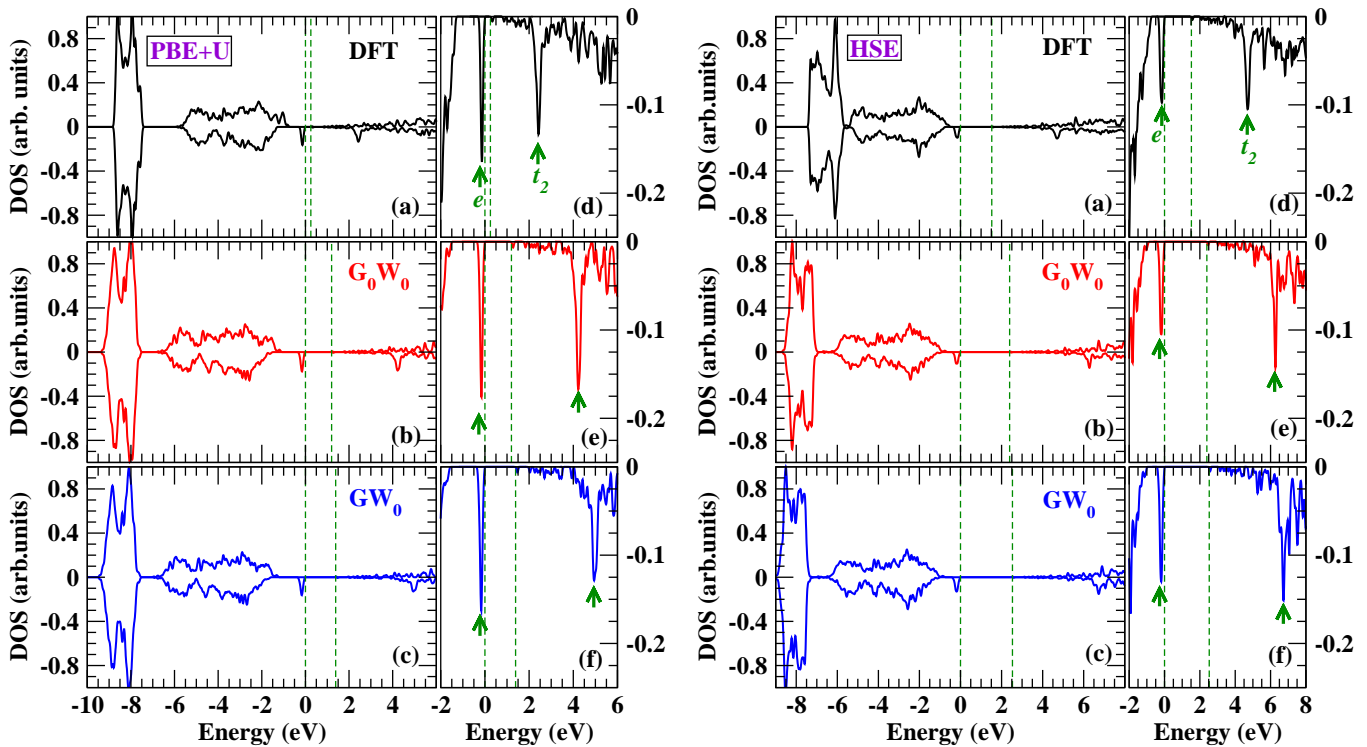}  
\caption{\label{SC-PBEUGW}Calculated density of states (DOS) in ZnO:Co
for a Co dopant concentration of 6.25 percent.~The DOS from two groundstate 
DFT starting points PBE+$U$ (left), HSE (right) and corresponding 
$G_0W_0$, $GW_0$ corrections on top of either are shown. Panels (d)-(f) 
in either case show a zoomed in view of the minority-spin DOS around the
Fermi energy. Green arrows indicate the positions of the Co derived minority-spin
$e$ and $t_2$ states. Lower and higher dashed lines indicate the positions 
of the Fermi energy and of the conduction band minimum respectively. The Fermi
energy in each case is aligned to 0 eV.} 
\end{figure*}
Perturbative $G_0W_0$ corrections on top of DFT starting 
wavefunctions lead to systematic improvements in the resulting quasi-particle spectrum.
We see from table~\ref{Tab1} that irrespective of the 
starting DFT XC functional, both E$_g$ and E$_{3d}$ from $G_0W_0$ are 
corrected towards the experimental values and including partial self-consisteny
through $GW_0$ further improves the agreement. Nevertheless, the value of 
E$_g$ is seen to depend upon the DFT starting point~\cite{Fuchs2007}. 
$G_0W_0$ on top of PBE (PBE+$U$) leads to a value of E$_g$ that is still underesimated
by $\sim$34\% ($\sim$24\%). In contrast, E$_g$ from HSE+$G_0W_0$ at 3.14 eV,  
is within $\sim$9\% of experiment. At the $GW_0$ level, E$_g$ is further increased 
relative to $G_0W_0$ and is within 22\% of experiment irrespective of starting DFT functional.
In particular,  E$_g$ from HSE+$GW_0$ at 3.31 eV matches well with experiment.
Similarly, $G_0W_0$ and $GW_0$ QP shifts generally tend to increase E$_{3d}$ 
relative to the DFT starting point and towards PES. However, the  PBE+$U$ starting point
which includes a large on-site $U_\mathrm{Zn}$=7 eV to being with, seems to be an exception.
Quasiparticle corrections in this case are seen to slightly reduce E$_{3d}$ from 6.98 eV in the
DFT groundstate to 6.63 eV in PBE+$U$+$GW_0$. Overall, the $GWA$ leads to
a $\sim$ 1-1.5 eV underestimation of the $3d$ band irrespective of the starting point 
as has been noted previously in the literature~\cite{Fuchs2007}.~Our results for E$_g$ and E$_{3d}$ in 
table~\ref{Tab1} are in good agreement with earlier benchmark calculations on 
zinc-blende ZnO~\cite{Shishkin2007,Fuchs2007} taking into account
that E$_g$ in wurtzite ZnO is expected to be $\sim$0.2 eV larger~\cite{Fuchs2007}.

\subsection{ZnO:Co supercell}
Co substituting Zn (Co$_\mathrm{Zn}$) in ZnO is formally in a Co$^{2+}$ oxidation 
state with 7 electrons in the occupied Co$_{3d}$ orbitals. The 
nearly tetrahedral crystal field around Co$_\mathrm{Zn}$, splits the Co$_{3d}$ states
into a set of lower $e$ and higher $t_2$ states. Furthermore, Co$_\mathrm{Zn}$
assumes a high-spin configuration with ($e^{\uparrow}$)$^2$~($t_2^{\uparrow}$)$^3$
majority spin and ($e^{\downarrow}$)$^2$~($t_2^{\downarrow}$)$^0$ minority
spin occupancies resulting in a local magnetic moment of 3$\mu$B per site.
\begin{table}[htb]
\caption{\label{Tab2}E$_{e^\downarrow}$, E$_{t_2^\downarrow}$ indicating the energy 
positions of Co $e^{\downarrow}$ and $t_2^{\downarrow}$ states relative to the host 
valence band top are presented for different levels of theory. Results from
supercell calculations both without and with an oxygen vacancy (V$_\mathrm{O}$) next to 
the Co are presented. E$_{t_2^\downarrow}$-E$_g$ indicates the position of the 
$t_2^{\downarrow}$ states relative to the conduction band minimum (CBM). The last column
gives the position of the $t_2^{\downarrow}$ states relative to the CBM if the 
latter is shifted rigidly to reproduce the experimental band-gap (E$^{exp}_g$)
while holding E$_{t_2^\downarrow}$ fixed.
}
\begin{ruledtabular}
\begin{tabular}{c c c c c}
name    &E$_{e^\downarrow}$  &  E$_{t_2^\downarrow}$ &  E$_{t_2^\downarrow}$-E$_g$&  E$_{t_2^\downarrow}$-E$_g^{exp}$ \\
\hline
& & ZnO:Co & & \\
\hline
PBE+$U$                    &  1.2        &  3.3       &   1.6     &  -0.1     \\
PBE+$U$+$G_0W_0$  &  1.3       & 5.2        &    2.4    &    1.8     \\
PBE+$U$+$GW_0$      &  1.4       & 6.1        &   3.0    &     2.6     \\
\hline
HSE                             & 0.6         &5.0        &2.7           & 1.6     \\
HSE+$G_0W_0$          & 0.7         &6.6        &3.3           & 3.2     \\
HSE+$GW_0$              & 0.8         &7.1        &3.6           & 3.7     \\
\hline
& &ZnO:Co + V$_\mathrm{O}$&&\\
\hline
PBE+$U$                     & 0.6          &2.5     &0.2            &-1.0     \\
PBE+$U$+$G_0W_0$  & 0.9     & 4.0    &0.9            &0.6       \\
PBE+$U$+$GW_0$ & 1.0  & 4.4  & 1.3 & 0.9 \\  
\end{tabular}
\end{ruledtabular}
\end{table}
First we briefly discuss the electronic structure of ZnO:Co obtained
from groundstate DFT calculations. LDA/GGA XC functionals yield a qualitatively 
incorrect groundstate for ZnO:Co~\cite{Das2008,Lany2008} by incorrectly placing the occupied 
$e^{\downarrow}$ states in resonance with the CBM of ZnO resulting in spurious 
charge transfer to the host and fractional occupation of the $e^{\downarrow}$ orbitals.
This is due to a combination of underestimating both the host band-gap and the 
binding energy of the Co$_{3d}$ states. $b$-LDA approaches that
either partially or fully rectify these shortcomings reproduce the correct occupancy
of the $e^{\downarrow}$ states~\cite{Das2008,Lany2008,Walsh2008,Patterson2006} and a magnetic
moment of 3$\mu$B per Co$_\mathrm{Zn}$. As a general feature common to the different $b$-LDA methods, 
the fully occupied majority spin Co$_{3d}$ states hybridize with the O$_{2p}$ valence band states 
of ZnO with some Co$_{3d}$ DOS at the top of the host VBM. Different approaches however differ
substantially at a quantiative level in their description of the minority spin Co$_{3d}$ states.

Considering the case of PBE+$U$ with $U_\mathrm{Zn}$=7 eV and $U_\mathrm{Co}$=3 eV,  $J_\mathrm{Co}$=1 eV,
we find that even though E$_g$ is still underesimated, the $e^{\downarrow}$ orbitals are correctly 
occupied by two electrons and are located approximately 1.2 eV above
the host VBM (see Fig.~\ref{SC-PBEUGW} and table~\ref{Tab2}). Meanwhile, the empty $t_2^{\downarrow}$ 
states are resonant in the conduction band
with an onset at roughly 1.6 eV above the CBM and 3.3 eV above the host VBM. Note however that the 
CBM is still too low in energy as E$_g\approx$1.6 eV. Within PBE+$U$ and related approaches, 
the positions of the minority spin $e^{\downarrow}$, $t_2^{\downarrow}$ states with respect to 
the host VBM (denoted by E$_{e^\downarrow}$, E$_{t_2^\downarrow}$ respectively) are largely 
determined by the choice of the parameter $U_\mathrm{Co}$. These quantities E$_{e^\downarrow}$, 
E$_{t_2^\downarrow}$ are insensitive to additional on-site corrections employed on Zn$_{4s}$ 
orbitals, within a DFT+$U$ approach to also rectify the host CBM position. Therefore if E$_g$ 
is restored to its full value of 3.44 eV within such a description~\cite{Walsh2008}, the empty $t_2^{\downarrow}$
states would be approximately resonant with the host CBM suggesting that they could be partially 
occupied at relatively small electron doping concentrations. 

The picture that emerges from the HSE functional while qualitatively similar to that of PBE+$U$ 
is quantiatively rather different. We find that the E$_{e^\downarrow}$ at $\sim$0.6 eV is
slightly smaller than in PBE+$U$ but E$_{t_2^\downarrow}$ is substantially larger 
at $\sim$5.0 eV. The inclusion of a fraction of  Fock-exchange generally pushes up unoccupied 
states higher in energy and so the increased value of E$_{t_2^\downarrow}$ is expected. A similar
result is found with other hybrid functionals~\cite{Patterson2006}.
With an E$_g$ value of 2.3 eV, the onset of the $t_2^{\downarrow}$ states is roughly 2.7 eV above 
the host CBM which renders partial occupancy of these states vitually impossible for reasonable
electron doping levels. Even assuming the full experimental value for E$_g$ by rigidly shifting the
Zn-$4s$ CBM higher in energy while keeping the Co-$3d$ states fixed leaves the $t_2$ states
about 1.6 eV above the CBM. Thus, starkly different implications emerge from PBE$+U$ and HSE for 
carrier mediated ferromagnetism in ZnO:Co. A natural question then arises as to which of these two 
$b$-LDA descriptions is closer to the quasi-particle picture. 

In figure~\ref{SC-PBEUGW} we also present DOS for ZnO:Co from $G_0W_0$ and $GW_0$ calculations
applied on top of PBE+$U$ and HSE. In general, the final QP spectrum both at the $G_0W_0$ and 
$GW_0$ levels 
depends to some extent on the DFT starting-point (SP). Note that in particular, perturbative 
$G_0W_0$ and $GW_0$ corrections applied on top of the qualitatively incorrect PBE groundstate 
of ZnO:Co (not shown),~do not lead to any improvement and are therefore of little interest. 
$b$-LDA groundstate SPs on the other hand lead to more systematic results. We see that the 
QP shift on E$_{e^\downarrow}$ is rather small irrespective of the SP.  
Accordingly E$_{e^\downarrow}$ occurs at 1.3 eV (1.4 eV) for $G_0W_0$ ($GW_0$) on top of PBE+$U$ 
and at 0.7 eV (0.8 eV) for $G_0W_0$ ($GW_0$) on top of HSE. In contrast, E$_{t_2^\downarrow}$ 
not only shows a larger QP shift but the shift is also invariably towards higher 
energies compared to the $b$-LDA SP. On top of PBE$+U$, $G_0W_0$ ($GW$) leads to an 
E$_{t_2^\downarrow}$ of 5.2 eV (6.1 eV) which places the onset of the empty $t_2^{\downarrow}$  
states $\sim$2.4 eV (3.0 eV) above the calculated host CBM. Thus the $t_2^{\downarrow}$ 
quasiparticle levels are predicted to be much higher in the conduction band than suggested by 
PBE$+U$ at the DFT level. Even assuming the full value of E$_g$ as above puts the $t_2^{\downarrow}$
states about 1.8 eV (2.6 eV) higher than the CBM in $G_0W_0$ ($GW_0$). Similarly, $G_0W_0$ ($GW_0$) on 
top of HSE yields an E$_{t_2^\downarrow}$ of 6.6 eV (7.1 eV) with the onset of the 
$t_2^{\downarrow}$  states 3.3 eV (3.6) eV above the corresponding calculated CBM. Based on these 
results for different DFT starting points, we estimate the onset of the $t_2^{\downarrow}$ states to
be 3.0-3.6 eV above the CBM of ZnO:Co. We find that the magnitude of the QP shifts relative to the 
DFT SP are smaller for HSE than PBE+$U$. The overall change in E$_{t_2^\downarrow}$ going from the 
DFT SP to $GW_0$ is 2.1 eV and 2.8 eV in HSE and PBE+$U$ respectively while the change in 
E$_{t_2^\downarrow}$ with respect to the CBM 
(see E$_{t_2^\downarrow}$-E$_g$ in table~\ref{Tab2}) also shows a similar trend at respectively 
0.9 eV and 1.4 eV in HSE and PBE+$U$.  A comprehensive study by Fuchs \textit{et. al}
~\cite{Fuchs2007}, indicated that for a wide range of materials, overall best agreement with
experimental spectra was obtained at the HSE+$G_0W_0$ level. At this level, the Co $t_2^{\downarrow}$
onset is predicted to be 3.3 eV above the CBM which we offer as a best compromise estimate. This precludes
the possibilty of FM interactions being mediated by partial occupancy of $t_2^{\downarrow}$ states
under electron doping in the absence of additional structural defects. 
\begin{figure}[htbp]
\includegraphics[width=0.5\textwidth,clip=true]{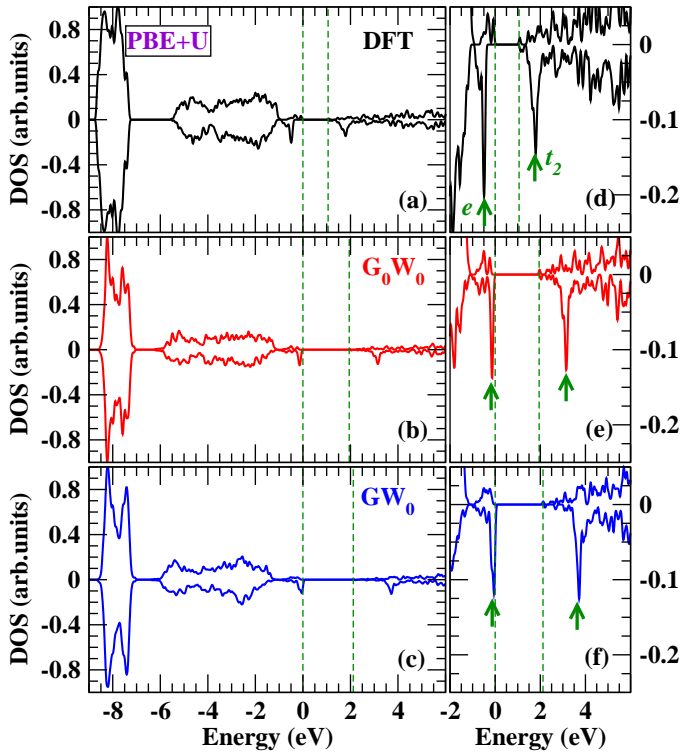}  
\caption{\label{VODOS}Calculated density of states (DOS) from a ZnO:Co
supercell containing a Co$_\mathrm{Zn}$ and oxygen vacancy (V$_\mathrm{O}$) pair.
~The DOS from groundstate 
DFT starting point PBE+$U$ (left) as well as corresponding 
$G_0W_0$, $GW_0$ corrections on top are shown. Panels (d)-(f) 
show a zoomed in view of the DOS around the
Fermi energy. Green arrows indicate the positions of the Co derived minority-spin
$e$ and $t_2$ states. Lower and higher dashed lines indicate the positions 
of the Fermi energy and of the conduction band minimum respectively. The Fermi
energy is aligned to 0 eV.} 
\end{figure}

Next we investigate the effect of low oxygen co-ordination around Co$_\mathrm{Zn}$ on the 
$e^{\downarrow}$ and $t_2^{\downarrow}$ QP energies. In an earlier work~\cite{Das2008} 
on ZnO:Co, based on a self-interaction corrected approach~\cite{ASIC2007,VPSIC2011},
we proposed that oxygen vacancies (V$_\mathrm{O}$) next to Co$_\mathrm{Zn}$ could lower
the energy of the $t_2^{\downarrow}$ states enough to make partial occupancy of these
states feasible at reasonable electron doping levels. In this context, we consider
one Co$_\mathrm{Zn}$+V$_\mathrm{O}$ pair in a nearest neighbour configuration within a
32 atom supercell of ZnO and calculate the QP energy levels for this system
at the $G_0W_0$ and $GW_0$ level based on a PBE+$U$ SP. The oxygen vacancy is
created by removing one out of the three O atoms co-ordinating the Co$_\mathrm{Zn}$
in the $ab$ plane of wurtzite ZnO. The calculated DOS for this system is
presented in figure~\ref{VODOS} and relevant energy levels are reported in 
table~\ref{Tab2}. We find that the effect of V$_\mathrm{O}$ next to Co$_\mathrm{Zn}$ is
to lower both E$_{e^\downarrow}$ and E$_{t_2^\downarrow}$. In fact the energy of
all the occupied Co-$3d$ manifold is lowered because of the smaller ligand field
the Co$_\mathrm{Zn}$ is now subject to~\cite{Das2008}. Relative to the case of an isolated
Co$_\mathrm{Zn}$, E$_{e^\downarrow}$ for the Co$_\mathrm{Zn}$+V$_\mathrm{O}$ pair is 
lower by 0.6 eV in PBE+$U$ and by 0.4 eV in both $G_0W_0$ and $GW_0$ on top of PBE+$U$.
The effect on E$_{t_2^\downarrow}$ is even larger as the crystal-field induced splitting
between $e^{\downarrow}$ and $t_2^{\downarrow}$ orbitals is also reduced. Thus at the PBE+$U$+$GW_0$ level
E$_{t_2^\downarrow}$ for a Co$_\mathrm{Zn}$+V$_\mathrm{O}$ pair is lower by almost 
1.7 eV relative to isolated Co$_\mathrm{Zn}$. The final alignment of the
$t_2^{\downarrow}$ states relative to the CBM, except in the case of PBE+$U$, 
is however still not favourable for driving carrier mediated FM. 
$G_0W_0$ ($GW_0$) places the onset of the $t_2^{\downarrow}$ states $\sim$0.9 eV (1.3 eV) 
above the CBM even for Co$_\mathrm{Zn}$+V$_\mathrm{O}$ pairs which sets a very
high electron-doping treshold~\cite{Lany2008} to achieve partial occupancy. Thus,
at the level of the $GWA$ considered in this work, the perspective that emerges
is decidedly more pessimistic for carrier mediated FM interactions between Co$_\mathrm{Zn}$
in ZnO:Co. We note that test calculations including self-consistency in the eigenvalues
in both $G$ and $W$ also produce qualitatively similar results with the QP shifts
being slightly larger in the same direction. These results are in line with the emerging 
consensus that ZnO:Co is not ferromagnetic in a conventional DMS sense~\cite{gamelin2008} 
but that the mechanisms responsible for the observed FM signatures are more exotic in
nature and perhaps confined to extended defects such as grain-boundaries in poly-crystalline
samples~\cite{hsu2007,Straumal2009}.

\section{CONCLUSION}
In conclusion, we investigated the quasiparticle (QP)
spectrum of Co doped wurtzite ZnO (ZnO:Co) within a $GW$ framework 
with a focus on the minority-spin $e^{\downarrow}$ and 
$t_2^{\downarrow}$ energy levels derived from Co
substituting a Zn site~(Co$_\mathrm{Zn}$). Single
shot $G_0W_0$ and partially self-consistent
$GW_0$ quasiparticle corrections were applied on top
of two different ground-state DFT starting points (SPs) based on 
the PBE$+U$ and HSE exchange-correlation functionals.
We find in general, the magnitude of the QP shifts to be smaller 
in the case of HSE compared to PBE$+U$. Irrespective
of the DFT SP and the level of $GW$ self-consistency, 
QP corrections are seen to shift the empty $t_2^{\downarrow}$ states
on Co$_\mathrm{Zn}$ higher in energy placing them roughly 3.3 eV
above the conduction band minimum (CBM) of ZnO:Co.~Low oxygen
co-ordination around the Co$_\mathrm{Zn}$ site lowers the QP energy
position of the $t_2^{\downarrow}$ states to around 1 eV above 
the CBM. Our results therefore suggest that  partial occupancy of 
the $t_2^{\downarrow}$ states by electron doping the host material 
is difficult to achieve making a conventional carrier mediated
mechanism for ferromagnetism in ZnO:Co less likely.

\section{ACKNOWLEDGEMENTS}
This work is funded by the Science Foundation of Ire-
land (???????????????) and by Isfahan University of
technology. Computational resources have been provided
by the Trinity Center for High Performance Computing.

\bibliography{ES}
\end{document}